\documentclass[pdflatex,sn-mathphys]{sn-jnl}
\usepackage{booktabs}

\jyear{2023}%

\theoremstyle{thmstyleone}%
\theoremstyle{thmstyletwo}%

\theoremstyle{thmstylethree}%

\begin{document}

\title[Nonlocal closures based on dispersion analysis]{Effects of operator nonlocality on closures for multicomponent reactive flows based on dispersion analysis}

\author*[1]{\fnm{Omkar B.} \sur{Shende}}\email{oshende@stanford.edu}

\author*[1]{\fnm{Ali} \sur{Mani}}\email{alimani@stanford.edu}

\affil*[1]{\orgdiv{Department of Mechanical Engineering}, \orgname{Stanford University}, \orgaddress{\city{Stanford}, \postcode{94305}, \state{CA}, \country{USA}}}

\abstract{Algebraic closure models with spatially nonlocal operators that are associated with both unresolved advective transport and nonlinear reaction terms in a Reynolds-averaged Navier-Stokes context are presented in this work. In particular, a system of two species subject to binary reaction and transport by advection and diffusion are examined by expanding upon analysis originally developed for binary reactions in the context of Taylor dispersion of scalars. This work extends model forms from weakly-nonlinear extensions of that dispersion theory and the role of nonlocality in the presence of reactions is studied and captured by analytic expressions. These expressions can be incorporated into an eddy diffusivity matrix that explicitly capture the influence of chemical kinetics and flow conditions on the closure operators and we demonstrate that the model form derived in a laminar context can be directly translated to an analogous setup in homogeneous isotropic turbulence, which has implications as a subgrid scale model. We show that this framework can improve prediction of mean quantities compared to previous purely local results, but does not fully close unresolved terms.}


\maketitle

\section{Introduction}\label{sec1}

Simulating systems of reacting scalars for the mean field values requires writing closures for not just a scalar flux term, as can be found in purely passive scalar transport, but also a scalar source term. The simplest possible setup for a reacting flow that captures this essential characteristic is a binary reaction setup. As such, consider the temporal evolution of such a system with two independent scalars with corresponding concentrations labeled as $C_1$ and $C_2$ in the presence of an imposed divergence-free velocity field, $U$, the governing equations for the scalars can be formulated as

\begin{equation}
    \frac{\partial C_i}{\partial t} + \nabla \cdot \left( UC_i \right) = D_m \nabla^2 C_i - AC_iC_{j\neq i},
    \label{eqn:DNS_full}
\end{equation}

\noindent where $A$ is a reaction coefficient and $D_m$ is a scalar diffusivity that is here assumed to be identical for both indexed scalars. When examining the evolution of the concentrations of scalars that do not affect such an incompressible flow, a one-way coupling between the fluid momentum and the scalars is assumed in writing the transport equations given by Equation (\ref{eqn:DNS_full}). This assumption is physically realizable in scenarios such as the dilute species limit, where the effects of heat release from the reaction are so small that neither the fluid density nor the reaction kinetics are significantly affected.

While this form of the equation incorporates all effects present in the system, in practice, the Reynolds-averaged analogue of Equation (\ref{eqn:DNS_full}), which can be written as 

\begin{equation}
    \frac{\partial \overline{C_i}}{\partial t} + \nabla \cdot \left( \overline{U}_{\,}\overline{C}_i \right) + \nabla \cdot \left( \overline{U'C_i'} \right) = D_m \nabla^2 \overline{C_i}
    - A\overline{C}_i\overline{C}_{j\neq i} - A\overline{C'_iC'_{j\neq i}},
    \label{eqn:RANS_full}
\end{equation}

\noindent is far less expensive to solve. This equation exists in same space as solutions to the Reynolds-averaged Navier-Stokes (RANS) equations, and we will therefore refer to it as a RANS scalar equation, even though we do not solve the RANS momentum equation herein. Primed terms represent fluctuations about the averaged quantities, which are denoted by over-bars. Finding models for the unclosed terms in Equation (\ref{eqn:RANS_full}), given by the scalar flux term, $\overline{U'C_i'}$, and a reaction source term, $\overline{C'_iC'_{j\neq i}}$, is essential to solving the full RANS problem and enabling faster predictions for computational problems in reacting flows and other fields where similar transport equations are solved, such as atmospheric pollution or oceanic biogeochemical cycles. \cite{seinfeld_2016}

In this work, the derivation of a model form from \cite{shende_2022} that uses weakly-nonlinear extension of dispersion analysis will be briefly recapitulated. In the following sections, we will first develop a model problem based on laminar flows that can capture dynamics neglected in that work for a linear reaction setup that will capture the effects of nonlocality. This involves the incorporation of information from a finite spatial kernel of support surrounding an evaluation point. The resulting nonlocal model form will be generalized using the reduced-order model (ROM) proposed in \cite{shende_2022} such that it can be applied to a binary reaction problem in a general turbulent flow. The outcome of such analysis is an algebraic model in the RANS context for nonlocal closures that clearly illustrates the role of chemical kinetics and nonreactive flow parameters on the magnitude of transport and reaction closures. The performance of this model form agains the local model form will then be discussed.

\section{Local formulation}

In this study, we seek to extend the work of \cite{shende_2022} to develop a model in the RANS context that can provide algebraic closures to the scalar evolution equations for a binary reactant setup. This is done in the spirit of dispersion analysis, as first demonstrated by Taylor in \cite{taylor_1953, taylor_1954} and since extended by many others to problems outside the context it was originally derived for \textit{cf.} \cite{aris_1956, lighthill_1966, mani_2021, prend_2021, daou_2018}. By considering the case of a parallel flow as an analytical prototype, \cite{shende_2022} extends Taylor's analysis to write local model forms for the case of dispersion of a system of binary scalars undergoing a reaction. This leads to scalar flux closures that can be written as

\begin{equation}
    \overline{u_i'C_j'} = -[D_{eff}]\left[\frac{\partial \overline{C}}{\partial x_i}\right], 
\end{equation}

\noindent where $D_{eff}$ is some generalized effective eddy diffusivity matrix and gradients of both reactants play a role. If considering a single scalar and invoking the standard gradient diffusion hypothesis, one could simply conclude that $D_{eff} = D^0$, where $D^0$ is a turbulent eddy diffusivity that is determined purely and fully by the flow. In the presence of reacting scalars, however, values of $D_{eff}$ differ from those observed in non-reactive cases and these differences can be explained though the use of a weakly-nonlinear extension of dispersion analysis. In particular, the effective eddy diffusivity involves the the superposition of two scalar gradients multiplied by a prefactor dependent on both scalar fields. The equations for the scalar flux are written as

\begin{equation}
\begin{bmatrix}
\overline{u' C_1'} \\
\overline{u' C_2'}
\end{bmatrix}
=-
\begin{bmatrix}
D_{11} & D_{12} \\
D_{21} & D_{22}
\end{bmatrix}
\frac{\partial}{\partial x}
\begin{bmatrix}
\overline{C_1} \\
\overline{C_2}
\end{bmatrix}
,
\label{eqn:ROM_transport}
\end{equation}{}   

\noindent where $D_{kl}$ represents the diffusivity coefficient associated with the flux of the $k$-th species due to a gradient in the $l$-th species. These coefficients are written explicitly as 
    
\begin{align}
\begin{split}
&D_{11} = \frac{D^0(1 + A\overline{C_1}\tau_{mix})}{1+A\tau_{mix}\overline{C_1}+A\tau_{mix}\overline{C_2}}  \\
&D_{12} = -\frac{D^0A\tau_{mix}\overline{C_1}}{1+A\tau_{mix}\overline{C_1}+A\tau_{mix}\overline{C_2}} \\
&D_{21} = -\frac{D^0A\tau_{mix}\overline{C_2}}{1+A\tau_{mix}\overline{C_1}+A\tau_{mix}\overline{C_2}}  \\
&D_{22} = \frac{D^0(1 + A\overline{C_2}\tau_{mix})}{1+A\tau_{mix}\overline{C_1}+A\tau_{mix}\overline{C_2}}.
\label{eqn:Ddefinitions}  
\end{split}
\end{align}  

Here, $\tau_{mix}$ denotes a characteristic mixing time associated with a particular turbulent flow. These two quantities, the mixing time and eddy diffusivity, are measured from an auxiliary, non-reactive flow. For example, the macroscopic forcing method (MFM) detailed in \cite{mani_2021, shirian_2022} allows for the diagnostic measurement of flow parameters by measuring the turbulent flow's response to analytic and exact imposed forcing.  

The most novel implication of the derivation approach used to obtain this model, however, is that it invokes an implied model for the fluctuating fields themselves. As a result, it is possible to determine the actual local form of the fluctuations and to close not only the scalar transport term, but also the unclosed reaction terms in the standard RANS equations. This closure expression can be written as 

\begin{equation}
    \overline{AC_1'C_2'} = \frac{A}{u_{rms}^2} \left( D_{11}\frac{\partial\overline{C_1}}{\partial x} + D_{12} \frac{\partial \overline{C_2}}{\partial x} \right) \left( D_{21}\frac{\partial\overline{C_1}}{\partial x} + D_{22} \frac{\partial \overline{C_2}}{\partial x} \right), 
    \label{eqn:ROM_scalar}
\end{equation}

\noindent where $u_{rms}$ is the measured root-mean-squared velocity field of the underlying flow. As such, we can fully close the scalar transport equation for both scalars using a single framework.

The basic weakly nonlinear model offers an alternative approach to perturbation expansions based on directly linearizing the equations performed by works like \cite{battiato_2011}. It recovers the scaling relationship between diffusivity and Damk\"ohler number derived in other numerical works and experiments, as in \cite{elperin_2014, elperin_2017, watanabe_2014, prend_2021}, and models based on analysis of single, linear reactions, such as in \cite{corrsin_1961} as explicated in \cite{peters_2000}. Unlike other pioneering work such as \cite{obrien_1971}, it does not require the scalar fields be purely homogeneous.

This provides a description of the local model that solves the reduced-degree-of-freedom system resulting from the RANS equations. The three algebraic components describe an entire model form, and require only information about the underlying flow, specifically $D^0$ and $u_{rms}$ and a known chemical kinetic parameter, $A$. 
 
The conclusion of \cite{shende_2022}, however, is that the purely local model form, even with higher-order corrections, is inadequate to fully capture the magnitudes of the closure terms.

\section{The model problem}

\begin{figure}[t!]
\centering
\includegraphics[height=350pt, angle=90]{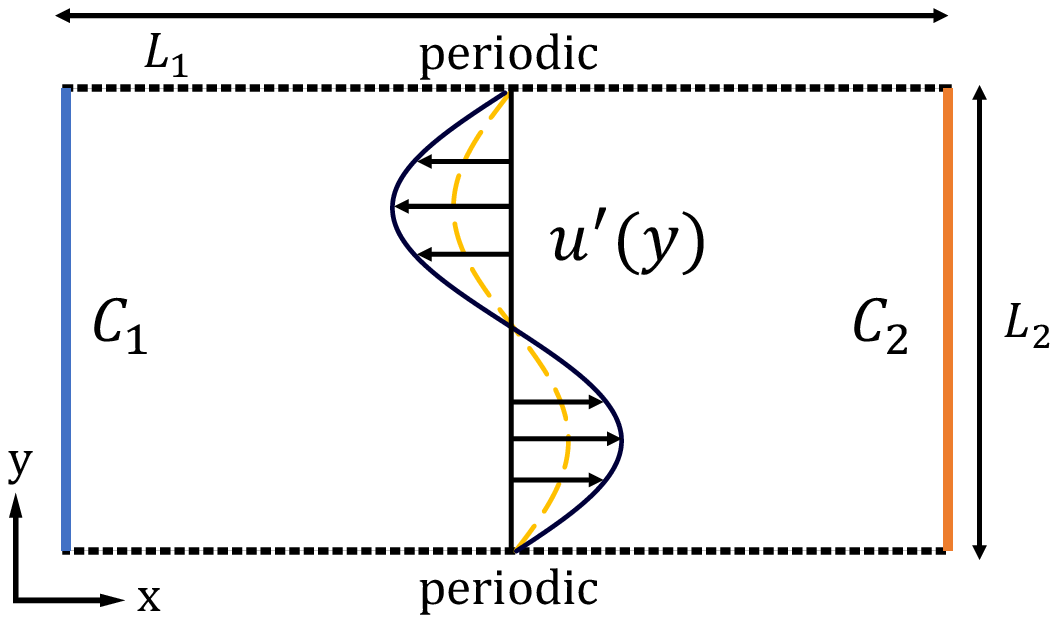}
     \caption{A schematic cartoon of the model problem, illustrating dimensions and boundary conditions. The flow is depicted in black as $u'(y)$, with a representation of a reaction front given at the mid-plane. Axial will refer to the $x$-direction, while spanwise will refer to the $y$-direction}
     \label{fig:cartoon}
\end{figure}

In order to develop insights into a model form that captures nonlocal closure terms in a binary mixture, we consider the following illustrative setup that can be treated semi-analytically: we impose a flow field that is parallel, steady, and two-dimensional and is contained in a domain comprised of an elongated box. We place periodic boundary conditions for the scalar fields on the two opposing elongated sides, as pictured in Figure (\ref{fig:cartoon}). By elongated, we mean that $L_1 >> L_2$.

In such a two-dimensional domain, an averaging operator for the RANS equations can be written as 

\begin{equation}
    \overline{f(x)} = \frac{1}{L_2} \int_{0}^{L_2} f(x,y) dy, 
\end{equation}

\noindent which differs from the pure ensemble-averaged version also considered by Reynolds-averaging. In this periodic domain, we assume a periodic structure for the flow-field with no imposed mean velocity. Therefore, $u(x,y) = \overline{u} + u' = U_0 sin(ky)$, where $k = 2\pi/L_2$. At $x = -L_1/2$ and $x = L_1/2$, the two domain boundaries, we prescribe Dirichlet conditions for the scalar concentrations, as pictured in Figure (\ref{fig:cartoon}).

In addition, we consider a reaction of the form $C_1 + C_2 \rightarrow C_3$, which is irreversible and fully activated, but is driven by a finite rate law of mass action. This full setup is simpler than a three-dimensional fully turbulent flow with a complex mechanism of reacting scalars, but this model problem nonetheless captures the essential competing physics of mixing, transport, and reactions that govern mean scalar concentrations in the more realistic case. As such, we can now formulate nonlocal closures with interpretable forms in this sandbox context.

In particular, past work in \cite{shende_2022} and work on a similar non-reactive problem in \cite{mani_2021} has shown that this setup allows for the formulation of solutions that translate directly from a laminar flow topology to a turbulent context, so we can be reasonably confident in the utility of these simplifications.

\subsection{Nonlocal formulation} 

To derive insight into the role of nonlocality on a reacting system, let us first examine the context of an irrerversible and fully activated linear reaction with a single scalar, as studied in works like \cite{corrsin_1961}. Here, to explain the nonlocal contributions to the macroscopic closure, we are following the basic steps of the MFM procedure that is fully explained in \cite{mani_2021}. Complete details can be found in that reference. 

For this particular setup, the concentration of one of the scalars, $C_2$, is held constant in the entire domain at unity and does not evolve, while the maximum value of $C_1$, at the left side of the domain, is held to be at least an order of magnitude less than that of $C_2$. As $C_2$ is a constant, we will denote the reaction rate constant as $A_L=AC_2$, giving the product of a standard binary species reaction coefficient times the concentration of the constant species. The fluctuations for $C_1$ are now governed by the transport equation

\begin{equation}
    \frac{\partial C_1'}{\partial t} + \frac{\partial (u'C_1')'}{\partial x_1}  + u'(y) \frac{\partial \overline{C_1}}{\partial x_1} = D_m \frac{\partial^2 C_1'}{\partial x_i \partial x_i}
    -  A_L C_1', 
\end{equation}

\noindent where $C_1 = \overline{C_1} + C_1'$ and so on. This equation is derived by finding the full transport equation for $C_1$ and subtracting from it the evolution equation for $\overline{C_1}$. Note that this equation is linear in the scalar concentration, and therefore can be described as evolving via a linear operator acting on the scalar field $C_1$. 

We further regard the term on the left-hand side that deals with fluctuations of fluctuations as negligible, following the assumptions of standard Taylor-type dispersion analysis. If we now add an arbitrary imposed forcing in time and the axial $x-$ dimension to the equation, we can write the transport equation as

\begin{equation}    
    \frac{\partial C'_1}{\partial t} + sin(x_2)\frac{\partial \overline{C_1}}{\partial x_1} = \frac{\partial^2 C'_1}{\partial x_2^2}
    + \frac{1}{Pe} \frac{\partial^2 C'_1}{\partial x_1^2} - A_L C'_1 + s(x_1,t), 
    \label{eqn:nl_linear_reac} \\
\end{equation}

\noindent where $s = exp(i\omega t + ikx_1)$ is an order one forcing term added to the right-hand side and $Pe = uL/D_m$ is a P\'eclet number. This forcing allows us to treat the linear operator as an input-output transfer function.

We can then also perform an expansion of the scalar concentration field as $C'_1 = \hat{c}(\omega, k;x_2) exp(i\omega t + ikx_1)$ as suggested by the imposed velocity field. This allows us to write a final governing equation of

\begin{equation}
    \left[ i\omega + \frac{k^2}{Pe} + A_L - \frac{\partial^2}{\partial x_2^2} + ik sin(x_2)  \right] \hat{c} = 1,
    \label{eqn:microscopiceqn}
\end{equation}

\noindent where $\omega$ is a temporal frequency and the bracketed terms represent a linear operator that acts on the scalar field. Note that we can now write this equation simply as $\hat{L}\hat{c}(\omega, k;x_2)=1$, where $\hat{L}$ denotes the full linear operator in its spectral form.

\begin{figure*}[t!]
\centering
\includegraphics[width=330pt]{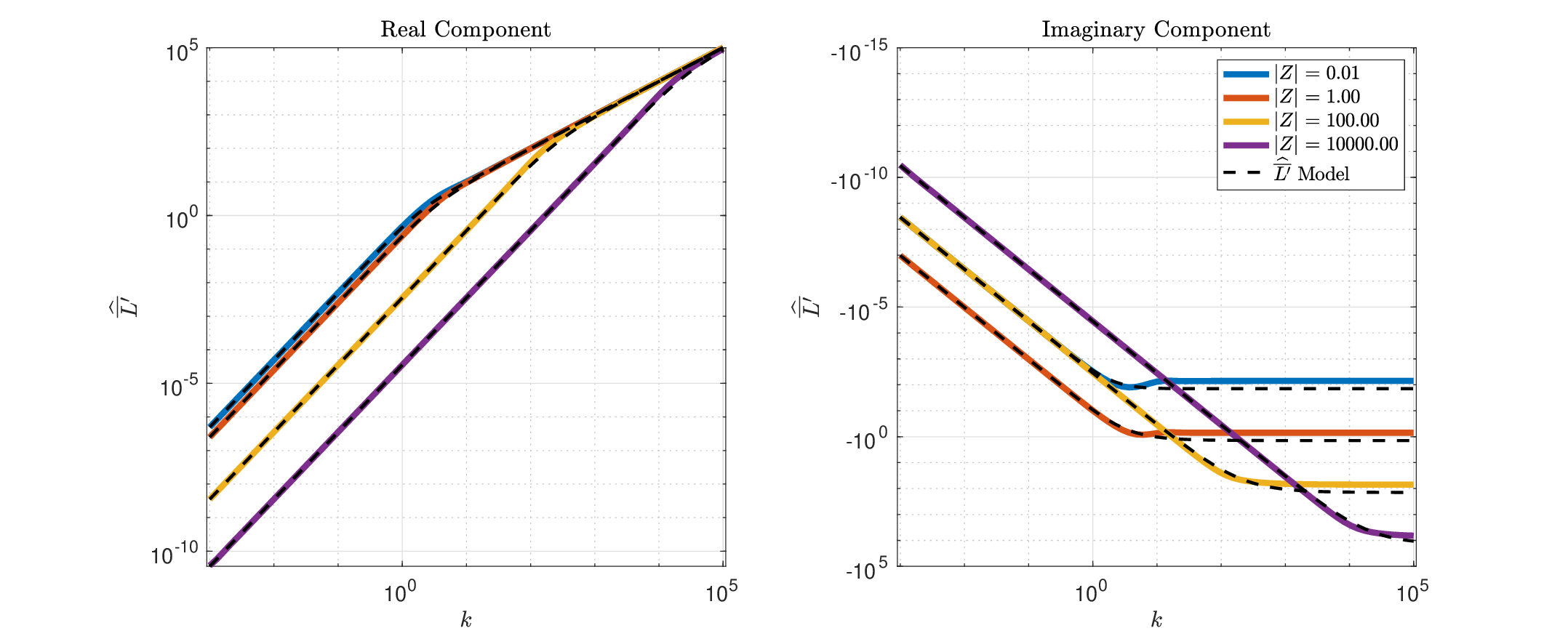}
\vspace{10 pt}
\caption{Plot of macroscopic operator as a function of the wavenumber for various magnitudes of the multi-physics coefficient with a fit of the proposed model of effective eddy diffusivity as given by Equation (\ref{eqn:Lhatbar}). In this case, $Z = \|Z\|e^{i\pi/4}$ and unsteady effects are included.}
\label{fig:nonlocal_deff1}
\vspace{20 pt}
\end{figure*}

\begin{figure*}[ht]
\centering
\includegraphics[width=330pt]{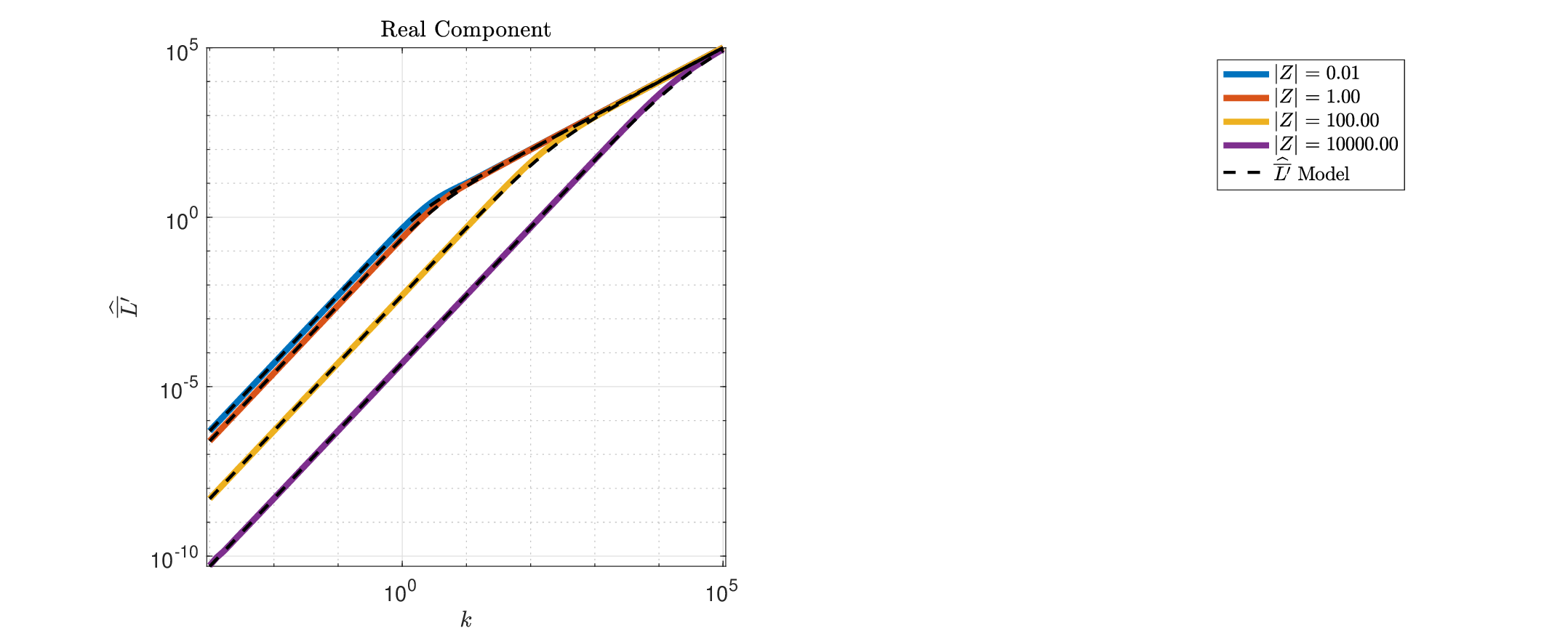}
\vspace{10 pt}
\caption{Plot of macroscopic operator as a function of the wavenumber for various magnitudes of the multi-physics coefficient with a fit of the proposed model of effective eddy diffusivity as given by Equation (\ref{eqn:Lhatbar}). In this case, $Z$ is purely real and we are in the steady limit. The imaginary portion in both model and MFM results is zero. }
\label{fig:nonlocal_deff2}
\end{figure*}

Thus far, we have followed the process of MFM as described in \cite{mani_2021} on a reactive system, and this has given us the exact operator in the microscopic, or un-averaged space. We note that we can also simplify this equation by rewriting it more generally as 

\begin{equation}
    \left[ Z - \frac{\partial^2}{\partial x_2^2} + ik sin(x_2) \right] \widehat{c}= 1,
\end{equation}

\noindent where $Z$, which we call a ``multi-physics coefficient,'' is a complex number that incorporates the effects of axial diffusion, chemical reaction, and unsteadiness. If we were to apply the over-barred averaging operator to this equation, only $Z$ and the advective term will remain, allowing us to find the operator that acts directly on $\overline{\widehat{c}}$, the averaged scalar field magnitude.

TO ALI: Averaging should commute with the expansion, so $\overline{\widehat{c}}$ = $\widehat{\overline{c}}$? The MFM paper switches between $\overline{\widehat{L}}$ and $\widehat{\overline{L}}$... 

We can solve this ordinary differential equation directly for each $\omega$ and $k$ and get $\widehat{c}$, the scalar magnitude, using appropriate boundary conditions. Now, after applying the averaging operator to the measured solution field, we have found the operator that acts on the full field in the spectral space. Thus, the macroscopic closure operator that captures purely unresolved effects in Fourier space can be computed by subtracting the mean field effects. We therefore invert Equation~(\ref{eqn:microscopiceqn}) and subtract off the closed portion of the operator so that we can write 

\begin{equation}
    \widehat{\overline{L'}} = (\widehat{\overline{c}})^{-1} - Z
\end{equation}

\noindent where the overbar represents averaging. This operator is now a function of the real spatial wavenumber, $k$, and $Z$, which can take complex values.

\subsection{The closure model form}

To explore the effects of the real and imaginary parts of the operator, we can consider $Z$ values that take the form of a scalar multiplied by $1+i$. We can plot the real and imaginary components of the macroscopic operator as in Figure (\ref{fig:nonlocal_deff1}) for this case, but if we consider purely the steady limit ($\omega = 0$), we can get Figure (\ref{fig:nonlocal_deff2}), which expresses the true steady macroscopic closure operator over a range of $Z$ values. In this limit, we see that the effect of increasing the reaction coefficient in Equation (8), the full governing equation, is to suppress the projection of perturbations to the macroscopic space.

While we have calculated the values for the macroscopic operator directly, the practical utility of this process is that one can now fit analytical curves to the actual calculated MFM operator. We do this by examining the asymptotic limits of said operator at zero and high $k$ and matching those limits for intermediate wavenumbers. One analytic expression that matches the limit values of $\widehat{\overline{L'}}$ is

\begin{equation}
    \widehat{\overline{L'}} = \left( (1+Z)^2 + k^2 \right)^{1/2} -1 - Z,
\end{equation}

\noindent which is a reasonable approximation to the true operator across decades of wavenumbers. We want to consider the steady limit, and a fit that approximates the analytic expression is

\begin{equation}
    \widehat{\overline{L'}} = \frac{k^2}{\sqrt{4(1+Z)^2 + k^2}}, 
\end{equation}

\noindent which allows us to consider the steady RANS problem. However, we want to modify this expression slightly in order to match previously explored model forms in the local linear limit as well as in the nonlocal, but nonreacting, limits. 

For a linear reaction problem, \cite{corrsin_1961} explicated a model form that should be recovered by this new generalized nonlocal operator. In the limit of non-reactive flow, we would also like to recover the non-reactive scalar transport model derived in \cite{mani_2021}. To do so, we can rewrite the operator derived here for the linear reaction setup as 

\begin{equation}
   \widehat{\overline{L'}}  = \frac{ \widehat{\overline{L'}}_{nr} }{1+Z \tau_{mix}} = \frac{ \frac{0.5 k^2}{\sqrt{1 + 0.25k^2}}}{1+\frac{Z}{\sqrt{1+0.25k^2}}},
   \label{eqn:Lhatbar}
\end{equation}

\noindent where the subscript $_{nr}$ denotes the non-reactive macroscopic operator. In the absence of reaction, the only term that survives is $\widehat{\overline{L'}}_{nr}$, which can be thought of as an effective eddy diffusivity multiplied with some gradient operators. This $D_{eff}$ matches the nonreactive but nonlocal scalar transport model derived in \cite{mani_2021}. 

This model for the eddy diffusivity is plotted in Figures (\ref{fig:nonlocal_deff1}-\ref{fig:nonlocal_deff2}) against the full measured MFM operator and we see good agreement. We further see that the role of the multi-physics coefficient is to suppress the magnitude of the macroscopic operator, which is captured in Equation (\ref{eqn:Lhatbar}). In the specific steady case where $Z = A_L$ and there is no axial diffusion, we recover the model form suggested by the local work of \cite{corrsin_1961} in the case of $k = 0$. 

\begin{figure}[t!]
\centering
\includegraphics[width=4in]{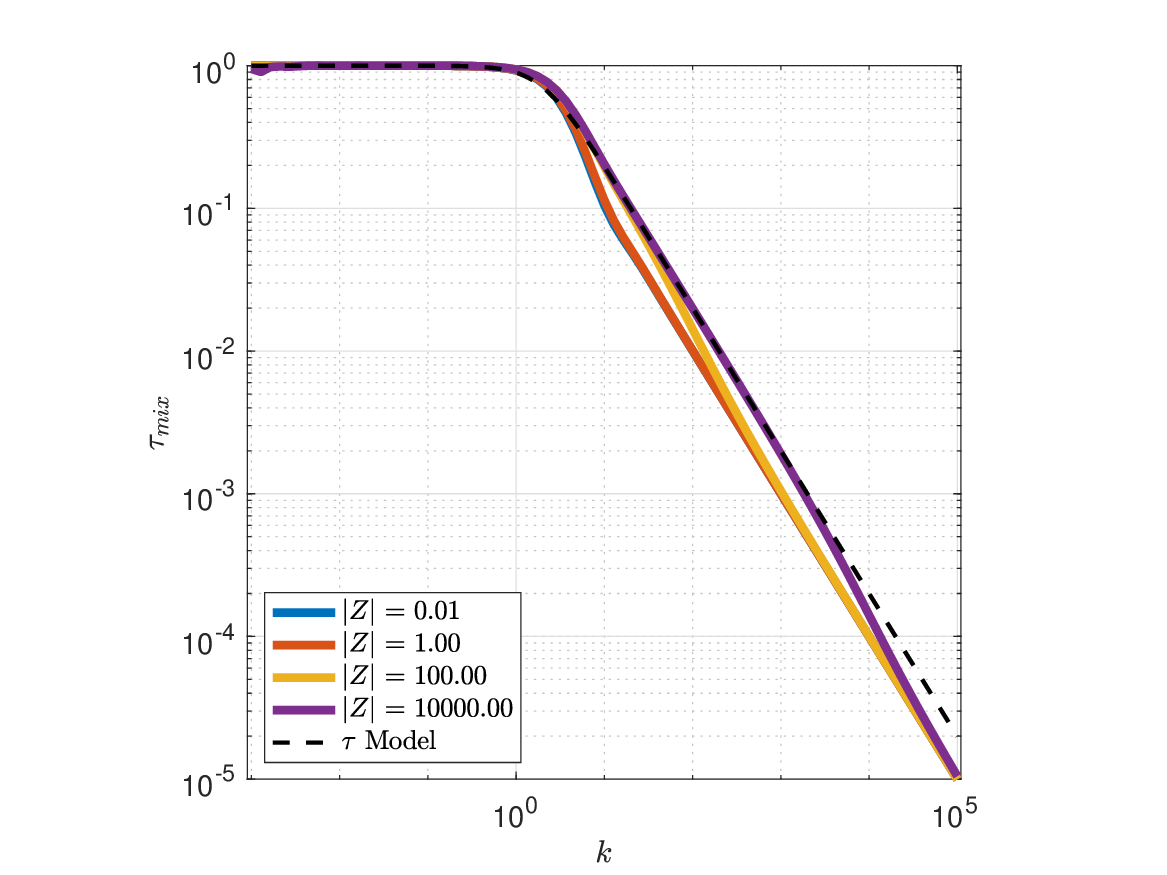}
\caption{Plot of mixing time as a function of the wavenumber for various magnitudes of the steady-state and real multi-physics coefficient with a fit of the proposed mixing time.}
\label{fig:tau}
\end{figure}

Furthermore, Equation (\ref{eqn:Lhatbar}) isolates a nonlocal form for the mixing time scale for this problem, 

\begin{equation}
    \tau = \left( 1+0.25k^2 \right)^{-1/2},
\end{equation}

\noindent and this assumed model form is plotted in Figure (\ref{fig:tau}). The colored lines represent the real measured mixing time as suggested by the middle expression in Equation (\ref{eqn:Lhatbar}) from a range of $Z = A_L$ values. We see the mixing time is highly scale-dependent, and is largely agnostic to the presence of a reaction. While not covered here, one would expect this mixing time to be sensitive to the steadiness of the underlying flow. 

From Figures (\ref{fig:nonlocal_deff1}-\ref{fig:nonlocal_deff2}), we can see that as the magnitude of reaction coefficient or the axial diffusivity increases, the magnitude of the nonlocal behavior of the closures rolls off.

\section{The turbulent problem}

We have thus far written an operator that works in Fourier space, but we must translate it into physical space to draw conclusions about its utility. We start by noting that each $ik$ term represents a spatial derivative, while each $i\omega$ denotes a temporal derivative. Now we can use the notation of $D^0$ and $l$, an eddy mixing length, from \cite{shirian_2022} to transform our operator derived from a laminar context into one suited for a fully turbulent flow. Such an operator for the linear reaction problem is given by

\begin{equation}
   D_{eff}  = \frac{D}{1+A_L \tau} \\ = \frac{D^0/\sqrt{\mathcal{I} - l^2 \frac{\partial^2}{\partial x_1^2}}}{\mathcal{I}+A_L \tau_{mix}/\sqrt{\mathcal{I}-l^2 \frac{\partial^2}{\partial x_1^2}}},
\end{equation}

\noindent where $\mathcal{I}$ represents an identity operator and which assumes that $Z = A_L$ only, as described in the previous section. We are concerned with the steady-state limit only here, and at the high $Pe$ limit, where we can neglect all terms in $Z$ except this scalar reaction coefficient.

It is important to note that while this model appears similar to a fractional step operator, it is derived analytically by considering the asymptotic limits of a governing equation solution. In addition, this operator is structurally similar to the work of \cite{corrsin_1961}, but is far more general and allows for definitions for mixing time and eddy diffusivity that incorporate nonlocality.

Based on the translation of the laminar model to a turbulent context, we can write that

\begin{equation}
   D  = \frac{D^0}{\sqrt{\mathcal{I}-l^2 \frac{\partial^2}{\partial x_1^2}}},
\end{equation}

and 

\begin{equation}
   \tau  = \frac{\tau_{mix}}{\sqrt{\mathcal{I}-l^2 \frac{\partial^2}{\partial x_1^2}}},
\end{equation}

\noindent using values for the eddy diffusivity, mixing time, and mixing length as defined and tabulated in \cite{shirian_2022}. In particular, a similar form of the nonlocal timescale is reported in \cite{shirian_phd}, which results from solving an explicit transport equation for the scalar flux terms akin to a Kramers-Moyal expansion.

In previous work, \cite{shende_2022} derived model forms for closures of the binary reaction problem involving $\tau_{mix}$ and $D^0$. With the observation that we can write independent expressions for eddy diffusivity and mixing time, an extension of those parameters to the binary reaction case is straightforward by adapting the model forms of Equation (\ref{eqn:Ddefinitions}) for the binary reaction problem. The newly stated nonlocal eddy diffusivity, $D$, and nonlocal mixing time, $\tau$, can be used in place of the previously used local eddy diffusivity, $D^0$ and the local mixing time, $\tau_{mix}$. This means, for example, we can write the first element of the diffusivity matrix as

\begin{equation}
D_{11} = \frac{D(1 + A\overline{C_1}\tau)}{1+A\tau \overline{C_1}+A\tau \overline{C_2}}.
\end{equation}  

Having conceived this model form, we seek to test its applicability for the binary reaction problem. In particular, we replace the steady, parallel flow examined in the previous section with three dimensional (3D) homogeneous isotropic turbulence (HIT) in an elongated domain. To generalize the model problem, the code of \cite{pouransari_2016} in an incompressible mode was adapted for this work to simulate HIT in a 3D domain of size $(2 \pi)^3$. The resulting flow field is periodically extended in the axial $x$-direction to generate a $20\pi\times2\pi\times2\pi$ computational domain for scalar transport. The scalar transport equations are solved with $C_1 = C_{ref}$ at $x = -L_1/2$ and $C_2 = C_{ref}$ at $x = L_1/2$. This provides a realistic reaction zone in the middle of the domain, far from the boundaries.

The velocity fields are solved on uniformly spaced structured meshes and the incompressible Navier-Stokes equations are solved with an added forcing of $Bu_i$ to sustain turbulence, where $B$ is a parameter described in \cite{rosales_2005}. The solver was run to statistical convergence as prescribed by \cite{shirian_2022} and we examine values at both $Re_{\lambda} = 26$ and $Re_{\lambda} = 40$, with $B = 0.2792$ for both Reynolds numbers. For $Re_{\lambda} = 26$, each box is a has $64^3$ points and the kinematic viscosity set to $\nu = 0.0263$, while for $Re_{\lambda} = 40$, each box is meshed with $128^3$ points with $\nu = 0.0111$. Five cases, with summary parameters given in Table \ref{tab:cases}, were studied and estimates of the turbulent statistical quantities, specifically $\epsilon$ and $u_{rms}$, in the table are adapted from \cite{shirian_2022}. The molecular diffusivity for both scalars is equal to the kinematic viscosity of the underlying fluid.

This setup allows adoption of the values for non-reactive eddy diffusivity. For all cases, we use values matching those in \cite{shende_2022}, which fall within error bounds reported by \cite{shirian_2022}. These values, chosen to match constant scalar flux values outside the reaction zone, for Case 1, for example, are $D^0 = 0.86u_{rms}^4/\epsilon = 0.96$, $l = 1.23u_{rms}^3/\epsilon = 1.41$, and $\tau_{mix} = 0.86u_{rms}^2/\epsilon = 1.02$, where $\epsilon = 0.79$ is turbulent kinetic energy dissipation rate and $u_{rms} = 0.97$ is the single-component root-mean-squared velocity. 

For scalar transport for Case 1, for example, we consider molecular diffusivity $D_m = 0.0263$, matching $\nu$, $C_{ref} = 1$ and a reaction coefficient of $A = 100$. This leads to $Pe \equiv D^0/D_m = 37$ and $Da \equiv A \tau_{mix} C_{ref} = 102$.

\begin{table}[ht]
\centering
\begin{tabular}{c|ccccc}
\hfil Case & \hfil 1 & \hfil 2 & \hfil 3 &  \hfil 4  & \hfil 5 \\ \hline
\hfil $Da$ & \hfil 104 & \hfil 52 & \hfil 26 & \hfil 138  & \hfil 276 \\
\hfil $Pe$ & \hfil 37 & \hfil 37 & \hfil 37 & \hfil 72  & \hfil 72 \\
\hfil $Re_\lambda$  & \hfil 26 & \hfil 26 & \hfil 26 & \hfil 40  & \hfil 40 \\
\hfil $u_{rms}$  & \hfil 0.97 & \hfil 0.97 & \hfil 0.97 & \hfil 0.905  & \hfil 0.905 \\
\hfil $\epsilon$  & \hfil 0.780 & \hfil 0.780 & \hfil 0.780 & \hfil 0.687  & \hfil 0.687 \\

\end{tabular}%
\caption{Summary parameters for the turbulent cases considered}
\label{tab:cases}
\end{table}

\subsection{\textit{A priori} analysis}

\begin{figure*}[b!]
\centering
\includegraphics[width=4in]{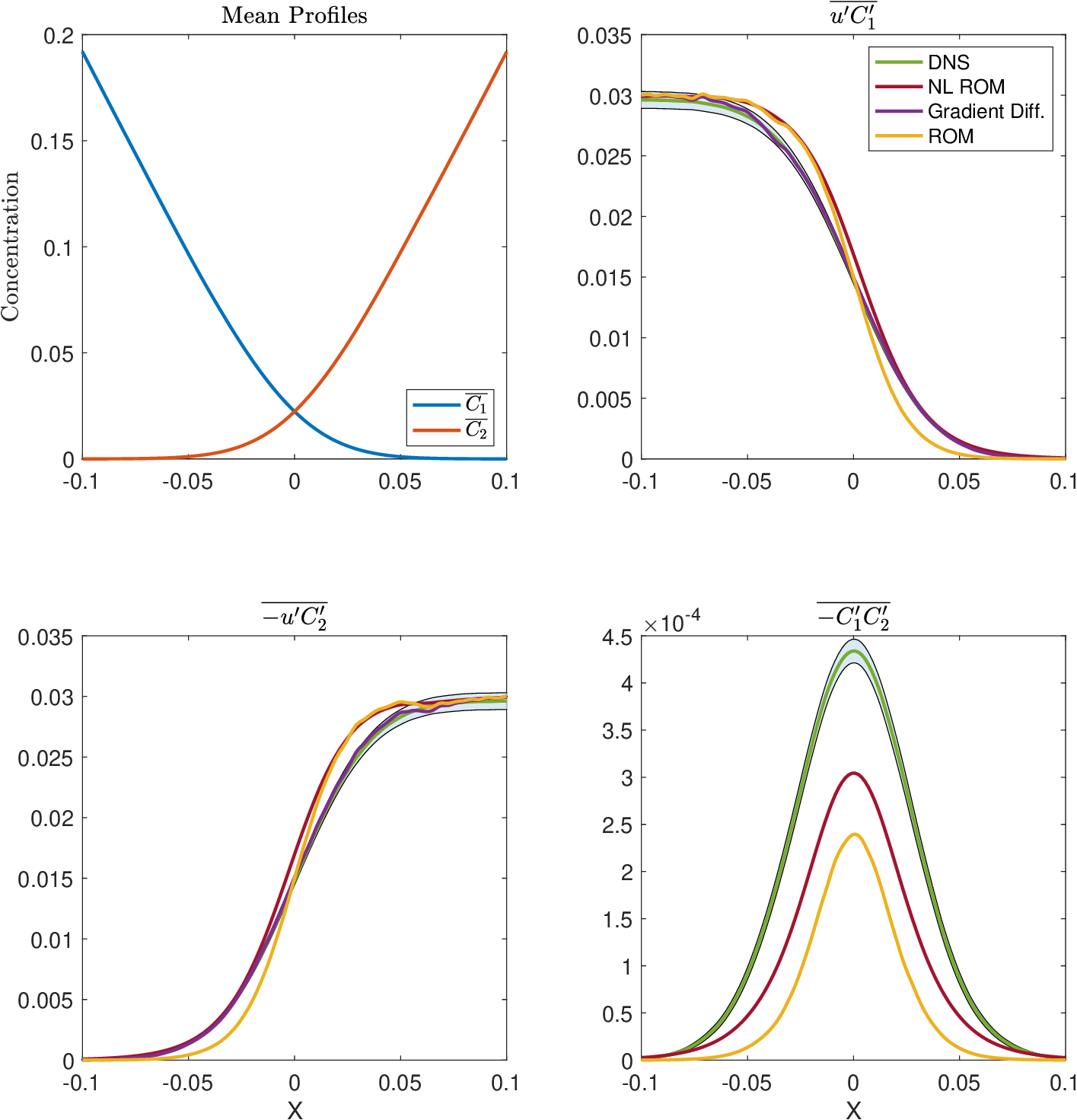}
\caption{Predictions of closures for the binary reaction problem subject to a turbulent flow. $X = x/L_1$ and DNS results are presented with 95\% confidence intervals accounting for statistical fluctuations.}
\label{fig:binary_apriori}
\end{figure*}

First, let us use \textit{a priori} analysis to examine the ensemble-averaged binary reaction problem in the defined turbulent context. In Figure (\ref{fig:binary_apriori}), we can see the plotted closures for Case 1, as well as the mean scalar concentrations. The nonlocal ROM recovers from some of the errors in capturing the true transport closures that are incurred by the local ROM model of \cite{shende_2022}. In particular, we see that the nonlocal model performs far better than the local model for the reaction closure term. 

For these specific closure terms, the standard Gradient Diffusion model appears to match the DNS data more closely than the local ROM and some regimes of the nonlocal ROM. However, an overall assessment that considers both reaction and transport closures reveals the advantage of the nonlocal ROM, as the Gradient Diffusion transport closure model offers no answers to the equally vital reaction closure question, and the local model does not accurately match the peak of the DNS reaction closure term. Quantifying this advantage can be accomplished by performing \textit{a posteriori} analysis and comparing one-dimensional model predictions against DNS data. 

While results are not omitted for the other cases, the conclusions are similar. As we are only using \textit{a priori} analysis as a proof-of-concept, we can now test the model forms directly by solving the full averaged equations.

\subsection{\textit{A posteriori} analysis}

In this section, having developed confidence in our nonlocal ROM, we can now solve Equation (\ref{eqn:RANS_full}) directly by invoking the closure models heretofore presented. The basic problem setup used in this section are identical to those for the \emph{a priori} analysis. 

The only exception is that for the RANS equation, instead of Dirichlet boundary conditions for the scalars, Neumann boundary conditions with slopes matching the DNS profiles are used. This is done because the DNS develops axial boundary layers near the Dirichlet conditions due to local outflow advection. For the DNS, we have ensured these artificial effects do not pollute the reaction zone results by ensuring the boundaries are far from the reaction zone. These boundary layers are absent in the RANS case as the mean velocity is zero. Appropriate matching of RANS solutions to the DNS ones should consider concentration profiles outside of these artificial boundary layers. We have done so by matching the slopes of the RANS concentration profiles to those of the DNS outside of the boundary layers, but far from the reaction zones.

\begin{figure}[t!]
\centering
\includegraphics[width=4in]{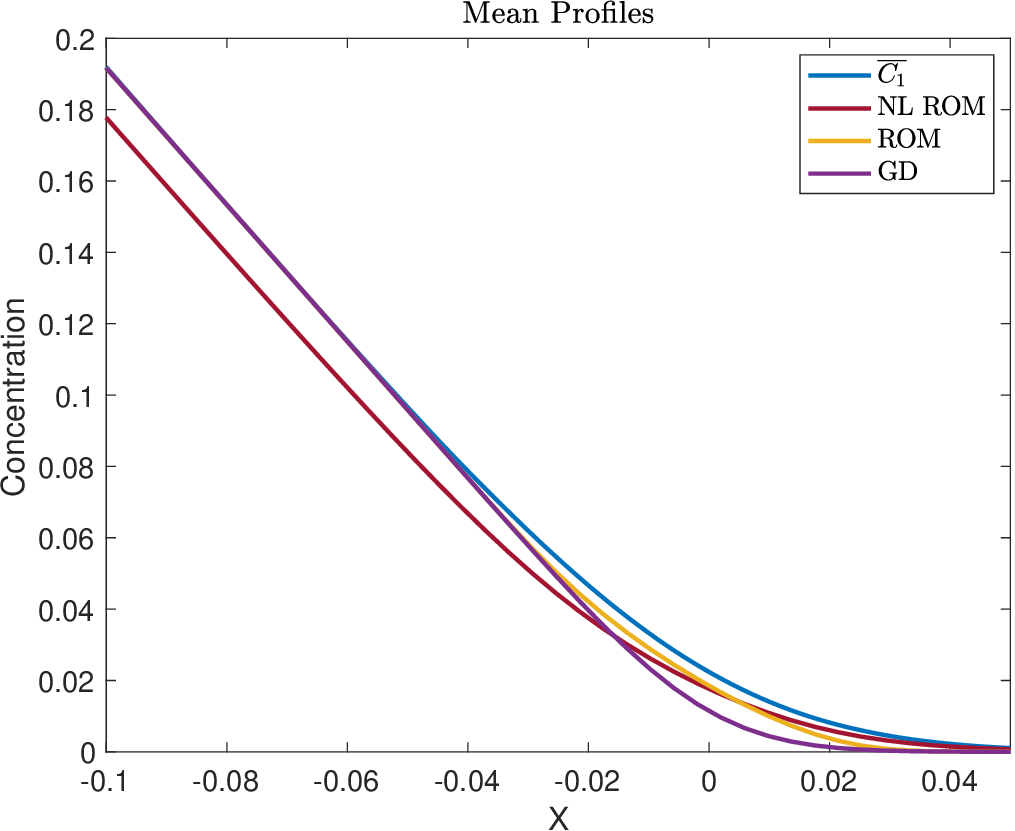}
\caption{Results of \emph{a posteriori} analysis, with DNS quantities indicated with overbars, the nonlocal formalism by \textbf{NL ROM}, and the standard local model with \textbf{ROM}, with $X = x/L_1$.}
\label{fig:binary_apost}
\end{figure}

In Figure (\ref{fig:binary_apost}), we can see the \textit{a posteriori} results for Case 1. In particular, we compare the predicted mean profile of $C_1$ to the results derived from the DNS described in the previous section. As the problem exhibits mirror symmetry in the ensemble-averaged sense about the midplane, results for $C_2$ are omitted as they provide the same insight. 

In Table (\ref{tab:errors}), we can see the maximum error incurred by each of the models in the center of the domain by comparing the predicted concentration for one of the scalars against the DNS measured values. 

\begin{table}[ht]
\centering
\scalebox{0.85}{
\begin{tabular}{c|ccccc}
\hfil Case & \hfil 1 & \hfil 2 & \hfil 3 & \hfil 4  & \hfil 5 \\ \hline
\hfil GD Max. Error     & \hfil $1.097\times10^{-2}$ & \hfil $9.192\times10^{-3}$ & \hfil $8.006\times10^{-3}$ & \hfil $1.035\times10^{-2}$  & \hfil $1.221\times10^{-2}$ \\
\hfil ROM Max. Error    & \hfil $4.588\times10^{-3}$ & \hfil $3.710\times10^{-3}$ & \hfil $3.131\times10^{-3}$ & \hfil $4.809\times10^{-3}$  & \hfil $5.878\times10^{-3}$ \\
\hfil NL ROM Max. Error & \hfil $1.426\times10^{-2}$ & \hfil $1.315\times10^{-2}$ & \hfil $9.412\times10^{-3}$ & \hfil $1.384\times10^{-2}$  & \hfil $1.575\times10^{-2}$ \\
\end{tabular}%
}

\caption{The maximum error in $C_1$ for the Gradient Diffusion (GD), Reduced Order Model (ROM), and Nonlocal Reduced Order Model (NL ROM) far from the boundaries as compared to the turbulent DNS results for all cases}
\label{tab:errors}
\end{table}

The local ROM outperforms the standard Gradient Diffusion model, and the nonlocal model also outperforms the GD model. However, for $X = x/L_1 \approx 0$, where the mean reaction zone, the ``flame,'' exists, Figure \ref{fig:binary_apost} shows that both models predict essentially the same value of concentration, while the nonlocal model better predicts the leakage of $C_1$ for $X = x/L_1 > 0$.

\subsection{Matrix Connections}

When considering the binary reaction case, it appears that the problem has changed significantly from the linear reaction case. However, it is possible to rewrite the model formulation of the eddy diffusivity for the binary reaction setup using matrix operators. In doing so, it takes on a form identical to the linear reaction case.

In Table \ref{tab:models}, we see in the first column the models for non-reactive flow, which match those proposed by \cite{mani_2021, shirian_2022}. The nonlocal model here contains the same operator form for $D$ as for the reactive cases. In the second column, we see the local and nonlocal models for the linear reaction problem. In this case, the effective eddy diffusivity is only modified by the presence of a background scalar, $C$. For the nonlocal problem, we recover the model derived using MFM analysis earlier in this work.

\begin{table}[ht]
\centering
\begin{tabular}{@{}llll@{}}
& No Reaction & Linear Reaction & Nonlinear Reaction \\[0.35cm]
Local Model & $D^0$ & $\frac{D^0}{1 + \tau A C}$ & $\frac{D^0}{\mathcal{I} + \tau A \textbf{C}}$ \\[0.35cm]
Nonlocal Model & $\frac{D^0}{{\sqrt{\mathcal{I}-l^2\nabla^2}}}$ & $\frac{\frac{D^0}{{\sqrt{\mathcal{I}-l^2\nabla^2}}}}{1 + \frac{\tau A C}{{\sqrt{\mathcal{I}-l^2\nabla^2}}}}$ & $\frac{\frac{D^0}{\sqrt{\mathcal{I}-l^2\nabla^2}}}{\mathcal{I} + \frac{\tau A \textbf{C}}{{\sqrt{\mathcal{I}-l^2\nabla^2}}}}$ \\[0.5cm]
\end{tabular}

\caption{A table showing all the models considered}
\label{tab:models}

\end{table}

In the two scalar case, denoted by the third column as involving a nonlinear reaction term, the scalar concentrations are given by a matrix, $\textbf{C}$, which can be written as 

\begin{equation}
\begin{bmatrix}
\textbf{C}
\end{bmatrix}
=
\begin{bmatrix}
C_{2} & C_{1} \\
C_{2} & C_{1}
\end{bmatrix}
,
\label{eqn:tensor_scalar}
\end{equation}{}  

\noindent which leads to a matrix expression in the denominator of the expression. This matrix, $\textbf{C}$, is obtained by linearizing the nonlinear reaction term, $AC_1C_2$, with respect to each of the two scalars in turn. If one evaluates the matrix expression and inverts the terms in the denominator, one recovers the expressions for the eddy diffusivity matrix from Equation \ref{eqn:ROM_transport}.

In short, this demonstrates that for pure scalar transport, there is a straightforward procedure to incorporate the effects of multiple scalars in both nonlocal and local model forms. Furthermore, we can frame the binary reaction diffusion solution as a linearization of the full problem.

\section{Conclusions}

In this work, we have introduced a framework to extend analysis created to study the dispersion of a single passive tracer to analysis of system of species undergoing binary reactions with nonlocal closures. The nonlocality for the binary reaction problem is derived from solving the corresponding linear reaction problem exactly to extract scale-dependent measures of eddy diffusivity and mixing time. The resulting model parameters and their incorporation into the model form of \cite{shende_2022} introduces closures to both advective flux and reaction terms with nonlinear interactions between the mean state of all species, even though the primary set of equations being solved undergo a linearization procedure. 

It is also salient to reiterate that while the capturing of nonlocality produces a model form that appear notionally similar to a fractional power operator, they are analytically explainable as fitting the high and low wavenumber limits of diffusion. In fact, \cite{liu_2021} uses analysis of moments of the full operator to write an alternative closed form solution of the expansion of the full operator without resorting to full tensorial operators.

In addition, while this work examines the steady-state limit of the nonlocal operator, the use of a generalized multi-physics coefficient in the model derivation allows the definition of an operator capable of modeling transient behaviour, including, for example, a time derivative in the algebraic operator.

However, a key takeaway is been that even for a notionally steady problem, i.e., the long-time response of a statistically stationary system, accounting for spatial nonlocality is insufficient to capture the behaviour of the mean scalar fields. Considering \textit{temporal} nonlocality, via a transport equation formulation that allows for the incorporation of history effects, must be the the logical missing piece of the puzzle. 

A future avenue of further inquiry is applying the proposed ROM to a Large-Eddy Simulation context as a subgrid-scale model, where mean space variables are replaced with filtered variables and quantities like $u_{rms}$ are evaluated at the grid scale. In particular, the Reynolds numbers considered herein are particularly relevant for the purposes of this modelling, where $x_1$ designates a flame-normal coordinate more generally. 

\bmhead{Acknowledgments}

Support for this work was provided by the National Science Foundation Graduate Research Fellowship Program under Grant No. 1656518, the Stanford Graduate Fellowships in Science and Engineering, and National Science Foundation Extreme Science and Engineering Discovery Environment resources under Grant No. CTS190057.

\bibliography{sn-bibliography}


\begin{thebibliography}{21}
\ifx \bisbn   \undefined \def \bisbn  #1{ISBN #1}\fi
\ifx \binits  \undefined \def \binits#1{#1}\fi
\ifx \bauthor  \undefined \def \bauthor#1{#1}\fi
\ifx \batitle  \undefined \def \batitle#1{#1}\fi
\ifx \bjtitle  \undefined \def \bjtitle#1{#1}\fi
\ifx \bvolume  \undefined \def \bvolume#1{\textbf{#1}}\fi
\ifx \byear  \undefined \def \byear#1{#1}\fi
\ifx \bissue  \undefined \def \bissue#1{#1}\fi
\ifx \bfpage  \undefined \def \bfpage#1{#1}\fi
\ifx \blpage  \undefined \def \blpage #1{#1}\fi
\ifx \burl  \undefined \def \burl#1{\textsf{#1}}\fi
\ifx \doiurl  \undefined \def \doiurl#1{\url{https://doi.org/#1}}\fi
\ifx \betal  \undefined \def \betal{\textit{et al.}}\fi
\ifx \binstitute  \undefined \def \binstitute#1{#1}\fi
\ifx \binstitutionaled  \undefined \def \binstitutionaled#1{#1}\fi
\ifx \bctitle  \undefined \def \bctitle#1{#1}\fi
\ifx \beditor  \undefined \def \beditor#1{#1}\fi
\ifx \bpublisher  \undefined \def \bpublisher#1{#1}\fi
\ifx \bbtitle  \undefined \def \bbtitle#1{#1}\fi
\ifx \bedition  \undefined \def \bedition#1{#1}\fi
\ifx \bseriesno  \undefined \def \bseriesno#1{#1}\fi
\ifx \blocation  \undefined \def \blocation#1{#1}\fi
\ifx \bsertitle  \undefined \def \bsertitle#1{#1}\fi
\ifx \bsnm \undefined \def \bsnm#1{#1}\fi
\ifx \bsuffix \undefined \def \bsuffix#1{#1}\fi
\ifx \bparticle \undefined \def \bparticle#1{#1}\fi
\ifx \barticle \undefined \def \barticle#1{#1}\fi
\bibcommenthead
\ifx \bconfdate \undefined \def \bconfdate #1{#1}\fi
\ifx \botherref \undefined \def \botherref #1{#1}\fi
\ifx \url \undefined \def \url#1{\textsf{#1}}\fi
\ifx \bchapter \undefined \def \bchapter#1{#1}\fi
\ifx \bbook \undefined \def \bbook#1{#1}\fi
\ifx \bcomment \undefined \def \bcomment#1{#1}\fi
\ifx \oauthor \undefined \def \oauthor#1{#1}\fi
\ifx \citeauthoryear \undefined \def \citeauthoryear#1{#1}\fi
\ifx \endbibitem  \undefined \def \endbibitem {}\fi
\ifx \bconflocation  \undefined \def \bconflocation#1{#1}\fi
\ifx \arxivurl  \undefined \def \arxivurl#1{\textsf{#1}}\fi
\csname PreBibitemsHook\endcsname

\bibitem{seinfeld_2016}
\begin{bbook}
\bauthor{\bsnm{Seinfeld}, \binits{J.H.}},
\bauthor{\bsnm{Pandis}, \binits{S.N.}}:
\bbtitle{Atmospheric Chemistry and Physics : from Air Pollution to Climate
  Change},
\bedition{Third edition.} edn.
\bpublisher{John Wiley and Sons},
\blocation{Hoboken, New Jersey}
(\byear{2016})
\end{bbook}
\endbibitem

\bibitem{shende_2022}
\begin{barticle}
\bauthor{\bsnm{Shende}, \binits{O.B.}},
\bauthor{\bsnm{Mani}, \binits{A.}}:
\batitle{Closures for multicomponent reacting flows based on dispersion
  analysis}.
\bjtitle{Phys. Rev. Fluids}
\bvolume{7},
\bfpage{093201}
(\byear{2022}).
\doiurl{10.1103/PhysRevFluids.7.093201}
\end{barticle}
\endbibitem

\bibitem{taylor_1953}
\begin{barticle}
\bauthor{\bsnm{Taylor}, \binits{G.I.}}:
\batitle{Dispersion of soluble matter in solvent flowing slowly through a
  tube}.
\bjtitle{Proceedings of the Royal Society of London. Series A. Mathematical and
  Physical Sciences}
\bvolume{219}(\bissue{1137}),
\bfpage{186}--\blpage{203}
(\byear{1953})
\end{barticle}
\endbibitem

\bibitem{taylor_1954}
\begin{barticle}
\bauthor{\bsnm{Taylor}, \binits{G.I.}}:
\batitle{The dispersion of matter in turbulent flow through a pipe}.
\bjtitle{Proceedings of the Royal Society of London. Series A. Mathematical and
  Physical Sciences}
\bvolume{223}(\bissue{1155}),
\bfpage{446}--\blpage{468}
(\byear{1954}).
\doiurl{10.1098/rspa.1954.0130}
\end{barticle}
\endbibitem

\bibitem{aris_1956}
\begin{barticle}
\bauthor{\bsnm{Aris}, \binits{R.}}:
\batitle{On the dispersion of a solute in a fluid flowing through a tube}.
\bjtitle{Proceedings of the Royal Society of London. Series A. Mathematical and
  Physical Sciences}
\bvolume{235}(\bissue{1200}),
\bfpage{67}--\blpage{77}
(\byear{1956})
\end{barticle}
\endbibitem

\bibitem{lighthill_1966}
\begin{barticle}
\bauthor{\bsnm{Lighthill}, \binits{M.J.}}:
\batitle{{Initial Development of Diffusion in Poiseuille Flow}}.
\bjtitle{IMA Journal of Applied Mathematics}
\bvolume{2}(\bissue{1}),
\bfpage{97}--\blpage{108}
(\byear{1966}).
\doiurl{10.1093/imamat/2.1.97}
\end{barticle}
\endbibitem

\bibitem{mani_2021}
\begin{barticle}
\bauthor{\bsnm{Mani}, \binits{A.}},
\bauthor{\bsnm{Park}, \binits{D.}}:
\batitle{Macroscopic forcing method: A tool for turbulence modeling and
  analysis of closures}.
\bjtitle{Phys. Rev. Fluids}
\bvolume{6},
\bfpage{054607}
(\byear{2021}).
\doiurl{10.1103/PhysRevFluids.6.054607}
\end{barticle}
\endbibitem

\bibitem{prend_2021}
\begin{barticle}
\bauthor{\bsnm{Prend}, \binits{C.J.}},
\bauthor{\bsnm{Flierl}, \binits{G.R.}},
\bauthor{\bsnm{Smith}, \binits{K.M.}},
\bauthor{\bsnm{Kaminski}, \binits{A.K.}}:
\batitle{Parameterizing eddy transport of biogeochemical tracers}.
\bjtitle{Geophysical Research Letters}
\bvolume{48}(\bissue{21}),
\bfpage{2021}--\blpage{094405}
(\byear{2021}).
\doiurl{10.1029/2021GL094405}
\end{barticle}
\endbibitem

\bibitem{daou_2018}
\begin{barticle}
\bauthor{\bsnm{Daou}, \binits{J.}},
\bauthor{\bsnm{Pearce}, \binits{P.}},
\bauthor{\bsnm{Al-Malki}, \binits{F.}}:
\batitle{Taylor dispersion in premixed combustion: Questions from turbulent
  combustion answered for laminar flames}.
\bjtitle{Phys. Rev. Fluids}
\bvolume{3},
\bfpage{023201}
(\byear{2018}).
\doiurl{10.1103/PhysRevFluids.3.023201}
\end{barticle}
\endbibitem

\bibitem{shirian_2022}
\begin{barticle}
\bauthor{\bsnm{Shirian}, \binits{Y.}},
\bauthor{\bsnm{Mani}, \binits{A.}}:
\batitle{Eddy diffusivity operator in homogeneous isotropic turbulence}.
\bjtitle{Phys. Rev. Fluids}
\bvolume{7},
\bfpage{052601}
(\byear{2022}).
\doiurl{10.1103/PhysRevFluids.7.L052601}
\end{barticle}
\endbibitem

\bibitem{battiato_2011}
\begin{barticle}
\bauthor{\bsnm{Battiato}, \binits{I.}},
\bauthor{\bsnm{Tartakovsky}, \binits{D.M.}}:
\batitle{Applicability regimes for macroscopic models of reactive transport in
  porous media}.
\bjtitle{Journal of Contaminant Hydrology}
\bvolume{120-121},
\bfpage{18}--\blpage{26}
(\byear{2011}).
\doiurl{10.1016/j.jconhyd.2010.05.005}
\end{barticle}
\endbibitem

\bibitem{elperin_2014}
\begin{barticle}
\bauthor{\bsnm{Elperin}, \binits{T.}},
\bauthor{\bsnm{Kleeorin}, \binits{N.}},
\bauthor{\bsnm{Liberman}, \binits{M.}},
\bauthor{\bsnm{Rogachevskii}, \binits{I.}}:
\batitle{Turbulent diffusion of chemically reacting gaseous admixtures}.
\bjtitle{Phys. Rev. E}
\bvolume{90},
\bfpage{053001}
(\byear{2014}).
\doiurl{10.1103/PhysRevE.90.053001}
\end{barticle}
\endbibitem

\bibitem{elperin_2017}
\begin{barticle}
\bauthor{\bsnm{Elperin}, \binits{T.}},
\bauthor{\bsnm{Kleeorin}, \binits{N.}},
\bauthor{\bsnm{Liberman}, \binits{M.}},
\bauthor{\bsnm{Lipatnikov}, \binits{A.N.}},
\bauthor{\bsnm{Rogachevskii}, \binits{I.}},
\bauthor{\bsnm{Yu}, \binits{R.}}:
\batitle{Turbulent diffusion of chemically reacting flows: Theory and numerical
  simulations}.
\bjtitle{Phys. Rev. E}
\bvolume{96},
\bfpage{053111}
(\byear{2017}).
\doiurl{10.1103/PhysRevE.96.053111}
\end{barticle}
\endbibitem

\bibitem{watanabe_2014}
\begin{barticle}
\bauthor{\bsnm{Watanabe}, \binits{T.}},
\bauthor{\bsnm{Sakai}, \binits{Y.}},
\bauthor{\bsnm{Nagata}, \binits{K.}},
\bauthor{\bsnm{Terashima}, \binits{O.}}:
\batitle{Turbulent schmidt number and eddy diffusivity change with a chemical
  reaction}.
\bjtitle{Journal of Fluid Mechanics}
\bvolume{754},
\bfpage{98}--\blpage{121}
(\byear{2014}).
\doiurl{10.1017/jfm.2014.387}
\end{barticle}
\endbibitem

\bibitem{corrsin_1961}
\begin{barticle}
\bauthor{\bsnm{Corrsin}, \binits{S.}}:
\batitle{The reactant concentration spectrum in turbulent mixing with a
  first-order reaction}.
\bjtitle{Journal of Fluid Mechanics}
\bvolume{11}(\bissue{3}),
\bfpage{407}--\blpage{416}
(\byear{1961}).
\doiurl{10.1017/S0022112061000615}
\end{barticle}
\endbibitem

\bibitem{peters_2000}
\begin{bbook}
\bauthor{\bsnm{Peters}, \binits{N.}}:
\bbtitle{Turbulent Combustion}.
\bsertitle{Cambridge Monographs on Mechanics}.
\bpublisher{Cambridge University Press}, \blocation{???}
(\byear{2000}).
\doiurl{10.1017/CBO9780511612701}
\end{bbook}
\endbibitem

\bibitem{obrien_1971}
\begin{barticle}
\bauthor{\bsnm{O'Brien}, \binits{E.E.}}:
\batitle{Turbulent mixing of two rapidly reacting chemical species}.
\bjtitle{The Physics of Fluids}
\bvolume{14}(\bissue{7}),
\bfpage{1326}--\blpage{1331}
(\byear{1971}).
\doiurl{10.1063/1.1693610}
\end{barticle}
\endbibitem

\bibitem{shirian_phd}
\begin{botherref}
\oauthor{\bsnm{Shirian}, \binits{Y.}}:
Application of macroscopic forcing method (mfm) for revealing turbulence
  closure model requirements.
PhD thesis,
Stanford University
(2022)
\end{botherref}
\endbibitem

\bibitem{pouransari_2016}
\begin{botherref}
\oauthor{\bsnm{Pouransari}, \binits{H.}},
\oauthor{\bsnm{Mortazavi}, \binits{M.}},
\oauthor{\bsnm{Mani}, \binits{A.}}:
Parallel variable-density particle-laden turbulence simulation
(2016).
\url{hadip@bitbucket.org/hadip/soleilmpi}
\end{botherref}
\endbibitem

\bibitem{rosales_2005}
\begin{barticle}
\bauthor{\bsnm{Rosales}, \binits{C.}},
\bauthor{\bsnm{Meneveau}, \binits{C.}}:
\batitle{Linear forcing in numerical simulations of isotropic turbulence:
  Physical space implementations and convergence properties}.
\bjtitle{Physics of Fluids}
\bvolume{17}(\bissue{9}),
\bfpage{095106}
(\byear{2005}).
\doiurl{10.1063/1.2047568}
\end{barticle}
\endbibitem

\bibitem{liu_2021}
\begin{botherref}
\oauthor{\bsnm{Liu}, \binits{J.}},
\oauthor{\bsnm{Williams}, \binits{H.}},
\oauthor{\bsnm{Mani}, \binits{A.}}:
A systematic approach for obtaining and modeling a nonlocal eddy diffusivity
(2021)
\end{botherref}
\endbibitem

\end{thebibliography}

\end{document}